\documentclass[a4paper,11pt]{article}
\usepackage{jcappub} 
\usepackage{lineno}

\usepackage{siunitx}
\usepackage{helvet}
\usepackage{amsmath}
\usepackage{amssymb}
\usepackage{geometry}
\usepackage{graphicx}
\usepackage{orcidlink}

\newcommand{\orcid}[1]{\orcidlink{#1}}

\title{\boldmath Investigating the Hubble Tension and $\sigma_8$
Discrepancy in $f(Q)$ Cosmology.}

\author[a,b,c]{Ziad Sakr,\orcid{0000-0002-4823-3757}}
\author[a]{Leonid Schey}
\affiliation[a]{Institute of Theoretical Physics, Philosophenweg 16, Heidelberg University, 69120, Heidelberg, Germany}
\affiliation[b]{IRAP, Université de Toulouse, CNRS, CNES, UPS, Toulouse, France}
\affiliation[c]{Faculty of Sciences, Universit\'e St Joseph; Beirut, Lebanon}

\emailAdd{sakr@thphys.uni-heidelberg.de}

\abstract{In this study, we incorporated a three-parameter family, of the metric incompatible modification of standard general relativity $f(Q)$ models into the Boltzmann code \texttt{MGCLASS} at both the background and perturbation levels, in order to conduct a Bayesian study employing probes that include the cosmic microwave background (CMB), baryon acoustic oscillations (BAO), weak lensing (WL), alone or its correlation with galaxy clustering (3$\times$2pt) and growth measurements $f \sigma_8$, for each submodel. Our analysis focused on the impact of the Hubble tension in $H_0$ and the discrepancy in $\sigma_8$ resulting from the inclusion of our model’s parameters, namely $M$, $\alpha$ and $\beta$. We find that none of the sub models, considered alone or combined, were able of alleviating the Hubble tension with only reducing it to 3\,$\sigma$ in the least constraining, highest degree of freedom case while we found that the $\sigma_8$ discrepancy, already strongly mitigated on WL linear scales, especially when we let all our model's parameters as free, appears again when considering the more constraining 3$\times$2pt probe. Among the parameters considered, we found that $\beta$, acting in scaling both the gravitational and the Hubble parameter, had the most impact in reducing the discrepancy, with data preferring far from $\Lambda$CDM alike values, before the combination with $f \sigma_8$ constrain it back to its general relativity values.}

\begin{document}
\maketitle
\flushbottom

\section{Introduction}
\label{sec:intro}

The improvement in the precision of observations, reveals that the predictions of $\Lambda$CDM are in tension with collected data \cite{Abdalla:2022yfr}. The arguably most notable one is the Hubble tension with a $5\sigma$ difference between the values of the Hubble constant ($H_0\sim73.0 \,\rm km. \, s^{-1} \,Mpc^{-1}$) inferred from the SH0ES team using Cepheid calibrated supernovae measurements  \cite{sh0es_h0} and the value of $H_0$ obtained from Planck's mission measurements of the cosmic microwave background (CMB) ($H_0\sim67.5 \, \rm km.  \, s^{-1} \,Mpc^{-1}$) \cite{Planck18.6}. Another mild one is the $\Omega_{m,0}$ - $\sigma_8$ \cite{Sakr:2018new,Sakr:2023hrl} or the related $S_8$ discrepancy with a $3\,\sigma$ difference in the values from Planck Mission 2018 (hereafter Plk18) release, $S_8\sim 0.832 $ \cite{Planck18.6} with respect to the weak lensing shear experiments from Kilo Degree Survey  (KiDS) bounds, $S_8\sim 0.76$) \cite{KiDS_shearConstraints} or Dark Energy Survey (DES), $S_8\sim 0.77$ \cite{DES:2021wwk}.

Alternatives beyond $\Lambda$CDM theories, have been deeply scrutinised, with some of them proving able to reduce the
tensions without however fully alleviating both at the same time, where moreover in some models, the resolution of one lead to the exacerbation of the other \cite{Akarsu:2024qiq,Sakr:2021jya,Gomez-Valent:2022bku,Hill:2020osr,Sakr:2023mao}. Among the above theories fall those which modify the long range gravitational interaction known as Modified
Gravity (MG) theories, based on an alteration of the Lagrangian, such as in Horndeski and beyond scalar-tensor theories \cite{Horndeski, beyond_Horndeski} or $f(R)$ theories \cite{mod_grav_nutshell,f(R)_theories} which are probably
the most prominent example of a class of modifications that directly tackle the realm
of geometry. In this work we investigate a certain model of the branch of $f(Q)$ theories \cite{FQ_review} which belong to the same class of geometrical modification, more specifically to the Symmetric Teleparallel Gravity (STG) \cite{jimenez2019geometrical,teleparallel}  in which gravity is attributed to its non-metricity and where $f(Q)$ is a general function of the non-metricity scalar $Q$.

Besides various cosmological studies to constrain $f(Q)$ limited however to the use of distance probes \cite{Oliveros:2023mwl,Koussour:2023hgl,Mandal:2023cag,Goswami:2023knh,Shi:2023kvu,Ferreira:2023awf,Gadbail:2023fjh,Maurya:2023fgy,Aggarwal:2022eae,Narawade:2022cgb,DAgostino:2022tdk,Lazkoz:2019sjl,Ayuso:2020dcu,Najera:2023wcw,Anagnostopoulos:2021ydo}, the interest in $f(Q)$-gravity grew in the last couple of years including works on black holes, wormholes, and modified stellar solutions, with the current state of the art reviewed in detail in \cite{FQ_review}.

In this work, we explore a class of $f(Q)$ models that draw from those proposed in \cite{DGPish} (see however other parametrisations in e.g. \cite{Gadbail:2024een,Khyllep:2022spx,Sokoliuk:2023ccw,Anagnostopoulos:2022gej,Esposito:2021ect}) where $f(Q)$ was parameterised as function of the parameter $M$ impacting only on the perturbation level, $\alpha$ modifying the Friedmann equation by scaling an additional Hubble parameter and $\beta$ that further scales the Gravitational constant. In \cite{DGPish}, the authors limited their investigation to the submodel ($\beta=1$, $M=0$ and $\alpha$ left free) using only geometrical distance probes, while \cite{can_fq_challenge} or \cite{RSD_fq_paper} added growth of structures probes in combination, but limited as well to the submodel ($\beta=1$, $\alpha=0$, $M\in\mathbb{R}$ free). 
 
Here we furthermore investigate all cases using  a large selection of probes on the geometrical or growth of structure level, an analysis enabled after we succeed in including our models in the Boltzmann code modification \texttt{MGCLASS II} \cite{MGCLASS2} and adapt likelihood pipelines within \texttt{MontePython} \cite{MoPy3} to perform various Markov Chain Monte Carlo (MCMC) simulations employing probes that comprise the cosmic microwave background (CMB), baryon acoustic oscillations (BAO), and weak lensing (WL) to examine whether the general model or even a submodel with less degrees of freedom is able to reduce the Hubble tension and $\sigma_8$ discrepancy. Very recently, a similar implementation, considering as well the most general case, was conducted by \cite{Goncalves:2024sem} in order to highlight the theoretical observable outputs in certain configurations that are equivalent to other common MG theories such as DGP models \cite{Dvali:2000hr}, without however going further to perform Bayesian studies in order to constrain the allowed space of parameters value with cosmological data nor study the impact on the aforementioned tensions. 

This paper is organised as follows: in Sect.~\ref{sect:theo} we review the STG theory and our $f(Q)$ adopted parameterisation. We describe the pipeline and data used in our analysis in Sect.~\ref{sect:method}. We present and discuss our results in Sect.~\ref{sect:results}, and conclude in Sect.~\ref{sect:conclusion}.

\section{Theoretical framework}\label{sect:theo}

In contrast to flat space, where vectors stay in a constant tangent space, the tangent space in General Relativity (GR) varies from point to point. Consequently, in order to make meaningful comparisons between vectors at different points, one equips the spacetime manifold having a metric $g_{\mu\nu}$ with an affine connection $\Gamma$, enabling the transportation of vectors to the same point, with a general formulation that can be split in three parts \cite{jimenez2019geometrical}
\begin{equation}
\label{general_connection}
    \Gamma^\alpha_{\ \mu\nu}=\{^\alpha_{\ \mu\nu}\}+K^\alpha_{\ \mu\nu}+L^\alpha_{\ \mu\nu}\ .
\end{equation}
The different parts are the Christoffel symbols 
\begin{equation}
    \{^\alpha_{\ \mu\nu}\}=\frac{1}{2}g^{\alpha\beta}(g_{\beta\nu,\mu}+g_{\mu\beta,\nu}-g_{\mu\nu,\beta})\ \mathrm{,}
\end{equation}
the contorsion tensor
\begin{equation}
    K^\alpha_{\ \mu\nu}=\frac{1}{2}T^\alpha_{\ \mu\nu}+T^{\ \ \alpha}_{(\mu \ \nu)}\ \mathrm{,}
\end{equation}
and the disformation tensor
\begin{equation}
    L^\alpha_{\ \mu\nu}=\frac{1}{2}Q^\alpha_{\ \mu\nu}-Q^{\ \ \alpha}_{(\mu \ \nu)}\ \mathrm{.}
\end{equation}
Einstein's original formulation of general relativity adheres the constraints of metric compatibility ($Q=0$) and vanishing torsion ($T=0$). Under these conditions, the connection simplifies to the Levi-Civita connection $\Gamma=\{^\alpha_{\ \mu\nu}\}$. 
However, in the so-called Symmetric Teleparallel Gravity (STG), we define the non-metricity tensor $Q$ with components \cite{jimenez2019geometrical}
\begin{equation}
    Q_{\alpha\mu\nu}=\nabla_\alpha g_{\mu\nu}\ .
\end{equation}
and a non-metricity scalar $Q$ by \cite{jimenez2019geometrical}
\begin{equation}
\label{Qscalar}
    Q =\frac{1}{4}Q_{\alpha\beta\mu}Q^{\alpha\beta\mu}-\frac{1}{2}Q_{\alpha\beta\mu}Q^{\beta\mu\alpha}-\frac{1}{4}Q_\alpha Q^\alpha+\frac{1}{2}Q_\alpha \Tilde{Q}^\alpha\ ,
\end{equation}
with the independent traces $Q_\mu=Q_{\mu \ \alpha}^{\ \alpha}$ and $\Tilde{Q}^\mu=Q_\alpha^{\ \mu\alpha}$.
From this scalar one can then introduce the action, with $R$ and $T = 0$ resulting in the same field equations as GR, the Symmetric Teleparallel Equivalent General Gravity (STEGR) \cite{jimenez2019geometrical}
\begin{equation}
\label{STEGR}
    S_{\mathrm{STEGR}}=\frac{1}{\kappa}\int \mathrm d^4x \ (\sqrt{-g} \, Q+\mathcal{L}_m\,
\end{equation}
with $\mathcal{L}_m$ the Lagrangian density of the matter fields and $\kappa=8\pi \,G /c^4$, where $G$ is the gravitational constant and $c$ the speed of light we set to 1.

    One of the common ways of modifying gravity based on the previous geometrical objects alone, without coupling to new fields is the $f(Q)$ extension in which we replace the non-metricity scalar $Q$ in Eq. \ref{STEGR} by a function $f(Q)$. The new action \cite{jimenez2019geometrical} 
    \begin{equation}
        S_{f(Q)}=\frac{1}{\kappa}\int \mathrm d^4x \ (\sqrt{-g}\,f(Q)+\mathcal{L}_m\ ,
    \end{equation}
    leads to generally modified Einstein equations depending on the specific model. More specifically, the field equations can be written in the form \cite{FQ_review, FQ_revisiting_cosmologies_review}: 
    \begin{equation}
        f_Q(Q)G_{\mu\nu}-\frac{1}{2}g_{\mu\nu}(f(Q)-f_Q(Q)f(Q))+2f_{QQ}(Q)P^\alpha_{\ \mu\nu}\partial Q_\alpha={T}_{\mu\nu}  \ ,
    \end{equation}
    where ${T}_{\mu\nu}$ is the stress energy tensor considered as having the form of a perfect fluid, i.e. ${T}_{\mu\nu} = {\rm diag} (-\rho,\,p,\,p,\,p)$, where $\rho$ is the energy density and $p$ the isotropic pressure, and 
    \begin{equation}
        f_Q=\frac{\partial f(Q)}{\partial Q}\ ,  \quad \quad    \quad  f_{QQ}=\frac{\partial^2 f(Q)}{\partial Q^2} \ ,
    \end{equation}
    and
    \begin{equation}
        P^\alpha_{\ \mu\nu}=-\frac{1}{4}Q^{\alpha}_{\ \mu\nu}+\frac{1}{2}Q^{\ \ \alpha}_{(\mu \ \nu)}+\frac{1}{4}g_{\mu\nu}Q^\alpha-\frac{1}{4}(g_{\mu\nu}\Tilde{Q}^\alpha+ \delta^\alpha_{\ (\mu} Q_{\nu)}) \ .
    \end{equation}
In order to derive the cosmology within this new framework, we express the FLRW metric 
\begin{equation}
    ds^2=g_{\mu\nu}\text{d} x^\mu \text{d} x^\nu=-g_{00}\,dt^2+a^2(t)\left[dx^2+dy^2+dz^2\right] \ ,
\end{equation}
 in $f(Q)$-geometry when one fixes the coincident gauge (implying $\Gamma^\alpha_{\ \mu\nu} = 0 $) \cite{cosmoinfq,teleparallel}. This gauge imposes constraints on the choice of coordinates, potentially leading to a modified element $g_{00}=-N(t)$ with the lapse function $N(t)$. Calculating the non-metricity scalar, we obtain $Q=6(H/N)^2$ for the modified metric. However, in this case, the $f(Q)$-action exhibits a time reparameterization symmetry, allowing us to freely choose $N(t)=1$.
Compared to GR, $f(Q)$-gravity leads to a modification of the Friedmann equations according to \cite{cosmoinfq}
\begin{align}
\label{mod_friedH}
    6f_QH^2-\frac{1}{2}f&=8\pi G\rho\ ,\\
    \label{mod_friedHdot}
    (12H^2f_{QQ}+f_Q)\Dot{H}&=-\frac{8\pi G}{2}(\rho + p)\ ,\\
\end{align}
with $f_Q=\frac{\partial f(Q)}{\partial Q}$, the dot refers here and throughout the text to cosmic time, and the model of $f(Q)$-gravity examined in this work is
\begin{equation*}
    f(Q)=\beta Q+\frac{\alpha}{2}\sqrt{Q}\log(Q/Q_{\mathrm{scale}})+M\sqrt{Q}\ .
\end{equation*}
From the modified Friedmann equations \ref{mod_friedH} and \ref{mod_friedHdot}, we derive the background equations
\begin{align}
\label{fqH2}
    H^2+\frac{\alpha}{6\beta}H&=\frac{8\pi G}{3\beta}\rho \ ,\\
    \dot H&=\frac{-4\pi G}{\beta}\sum(1+\omega_i)\rho_i\left(1-\frac{\alpha}{\sqrt{\alpha^2+64\pi G\beta\rho_{\mathrm{tot}}}}\right) \ .
\end{align}
Eq. \ref{fqH2} is an ordinary quadratic equation and has the positive-branch solution
\begin{equation}
\label{fqH}
    H=\frac{\sqrt{6}}{12\beta}\left(\sqrt{\alpha^2+64\pi G \beta \rho_{\mathrm{tot}}}-\alpha\right) \ . 
\end{equation}
In our model we include a certain amount of dark energy $\Omega_\Lambda=\rho_{\Lambda}/\rho_{\mathrm{crit}}$ in form of a cosmological constant such that the constrain from Eq. \ref{fqH2}
\begin{equation}
\label{lambda_fq}
    \Omega_\Lambda=\frac{3\beta}{8\pi G\rho_{\mathrm{crit}}}({H_0}^2+\frac{\alpha}{6\beta}H_0)-\left(\Omega_{\mathrm{m}}+\Omega_{\mathrm{r}}\right) \ ,
\end{equation}
is fixed for every suitable combination of $\alpha$, $\beta$, $\Omega_{\mathrm{m/r}}$, and $H_0$, with the critical density $\rho_{\mathrm{crit}}=3{H_0}^2/8\pi G$ \cite{amendola_tsujikawa_2010} and $\Omega_{\mathrm{m/r}}$ representing cosmic dust (non-relativistic matter) and radiation, respectively.

Our model consists of three different terms, each coupled to a control parameter that provides the freedom to adjust their individual impact. Their motivation and implications are the following:
\begin{itemize}
    \item Linear term ($Q$):
    The linear term alone corresponds to STEGR. As GR is still the widely accepted and fits most data very well, we include it and make rather small deviations from STEGR with the other parts of the model. Regarding the term on its own, the control parameter $\beta$ has the same impact as both a change in Newton's constant and in the additional Hubble parameter.
    \item Square-root term ($\sqrt{Q}$):
    The square-root term is a rather common added term in $f(Q)$-models (see e.g. \cite{can_fq_challenge, RSD_fq_paper, FQ_signatures, FQ_M_sirens, FQ_energybounds}). The reason for this is that the modified Friedmann equations remain unchanged by its addition. Hence, the square-root term gives the possibility to make changes on perturbation level while fixing the background. We name its control parameter $M$ following literature conventions.
    
    \item Logarithmic term [$\sqrt{Q}\log(Q/Q_{\mathrm{scale}})$]:
    The logarithmic term is specifically designed to lead to the linear extension in Eq.~\ref{fqH} \cite{DGPish} similarly to DPG cosmologies \cite{DGP_main, DGP_1}. The divisor in the logarithm's argument, $Q_{\mathrm{scale}}$, is in contrast to $\alpha$ no free parameter and it is fixed to 1 ${\rm Mpc}^{-2}$. We do not lose any generality fixing $Q_{\mathrm{scale}}$, because splitting the logarithm gives a term of $-\alpha\log(Q_{\mathrm{scale}})\sqrt{Q}$ which is also proportional to $\sqrt{Q}$, making $\alpha\log(Q_{\mathrm{scale}})$ and $M$ degenerate. The case $\alpha<0$ leads to a self-accelerated universe even if its only matter content would not produce acceleration on its own \cite{DGPish}. The opposite is given for $\alpha>0$ \cite{DGPish}. We followed the latter option when varying $\alpha$ in our MCMC exploration.
\end{itemize}
The parameters $\alpha$ and $M$ are bound to unit $\mathrm{length}^{-1}$. For the investigation of their best-fit values and the comparison to literature, we redefine both quantities in a dimensionless way as
\begin{align}
    \alpha &\rightarrow \alpha/H_0\ ,\\
    M &\rightarrow M/H_0\ .
\end{align}
The perturbed metric components in Newtonian gauge at linear order can be written as \cite{dodelson2020modern}
\begin{align}
    g_{00}(\Vec{x},t)&=-1-2\Psi(\Vec{x},t)\ ,\\
    g_{0i}(\Vec{x},t)&=0\ ,\\
    g_{ij}(\Vec{x},t)&=a^2(t)\delta_{ij}[1-2\Phi(\Vec{x},t)]\ ,
\end{align}
with the metric perturbations $\Psi$ and $\Phi$, in the quasi-static limit, deep inside the Hubble radius, according to \cite{cosmoinfq, FQ_signatures}, related to $\delta=\delta\rho / \rho$ the density contrast by the Fourier transformed Poisson equation in $f(Q)$ theory 
\begin{align}
    \label{mod_poisson}
    \Psi&=\frac{4\pi Ga^2\rho\delta}{f_Qk^2} \ ,\\
    \label{mod_potentialequality}
    \Psi &= \Phi \ .    
\end{align}
The factor in the denominator of the modified Poisson equation due to the model is
\begin{equation}
\label{f_Q_model}
    f_Q=\beta+\frac{\alpha}{2}\frac{1}{\sqrt{Q}}+\frac{\alpha}{4}\frac{\log(Q/Q_{\mathrm{scale}})}{\sqrt{Q}}+\frac{M}{2}\frac{1}{\sqrt{Q}}\ .
\end{equation}

\section{Analysis pipeline and datasets}\label{sect:method}

We implemented our gravity model in \texttt{MGCLASS}. The implementation uses the structure of \texttt{MGCLASS} which encapsulate the impact of such theories on the perturbation potentials $\Phi$ and $\Psi$ parameterized by two functions $\mu$ and $\eta$ encoding the possible deviations from GR with \cite{MGCLASS2}
\begin{align}
     \Psi_{\mathrm{MG}}(a,\Vec{k})&= \Psi_{\mathrm{GR}}(a,\Vec{k})\times\mu(a,k)\ ,\\
     \Phi_{\mathrm{MG}}(a,\Vec{k})&= \Psi_{\mathrm{MG}}(a,\Vec{k})\times\eta(a,k)\ .
\end{align}
In our case, according to Eq. \ref{mod_poisson} and \ref{mod_potentialequality}, reduce to
\begin{align}
    \mu&={f_Q}^{-1}\ ,\\
    \eta&=1\ .
\end{align}
Since the code also relies on the derivative of $\mu$ and $\eta$ with respect to time $\tau$, we have
\begin{align}
    \frac{\partial\mu}{\partial\tau} &=-\frac{a\dot f_Q}{{f_Q}^2}\ ,\\
    \dot f_Q&=-\frac{1}{4}\frac{\dot Q}{\sqrt{Q}^3}\left[\frac{\alpha\log Q}{2}+M\right]\ ,\\
    \dot Q&= 12H\dot H\ ,\\
    \frac{\partial\eta}{\partial\tau}&=0\ .
\end{align}
Our code and the implementation of our $f(Q)$-cosmology has been tested in various ways:
\begin{enumerate}
    \item Our model in its full generality is not encompassed by the theories already included in \texttt{MGCLASS}, but the sole linear term for $\beta\in\mathbb{R}$, $\alpha=0$, and $M=0$ could be mapped to the Lagrangian action of $f(R,\phi,X)$ with $F(\phi)=2\beta$, $X=0$, and $U(\phi)=0$ in the case where we choose the \texttt{bckg} option corresponding to a special feature in \texttt{MGCLASS} where the potential MG couplings impact as well the background. In this scenario the background and perturbation equations of $f(R,\phi,X)$-theories and those of our $f(Q)$-model are the same and therefore \texttt{MGCLASS} gave as expected the same results for various positive values of $\beta$ independent if either of the theory's equations used. 
    
    \item We plotted the angular diameter distance $d_\mathrm{A}$ given by \cite{amendola_tsujikawa_2010}
    \begin{equation}
        d_\mathrm{A}=(1+z)^{-1}\int^z_0\frac{\text{d} z'}{H(z')}\ ,
    \end{equation}
    to see whether the individual parameters have the correct impact. According to Eq. \ref{fqH} it should not be changed for $M$, it should be stretched by $\sqrt{\beta}$ for $\beta$, and in a more qualitative matter, it should be slightly above the GR curve for a small positive value of $\alpha$ as it is shown in Fig.~\ref{angdiadis}. Furthermore, the results agree with the predicted behavior of another distance quantity $H(z)$ drawn in \cite{Goncalves:2024sem}.
    
    \begin{figure*}[t]
    \centering
    \includegraphics[width=0.49\textwidth, height=0.3\textwidth]{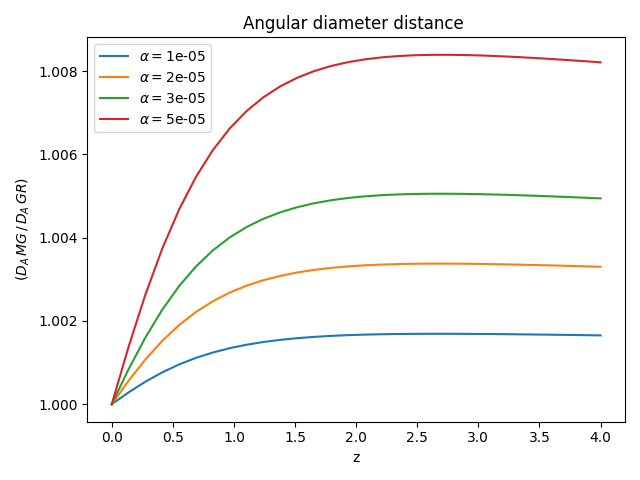}
    \includegraphics[width=0.49\textwidth, height=0.3\textwidth]{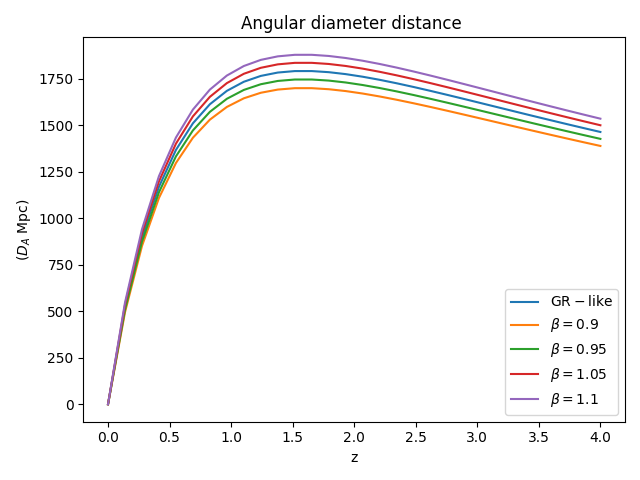}
    \caption[Relative and angular diameter distance $D_\mathrm{A}$.]{Left panel: Relative angular diameter distance $D_\mathrm{A}$ of our model compared to GR for different values of $\alpha$. Right panel: Angular diameter distance $D_\mathrm{A}$ of our model for different values of $\beta$.}
    \label{angdiadis}
    \end{figure*}

          \begin{figure*}[t]
    \centering
    \includegraphics[width=0.35\textwidth, height=0.3\textwidth]{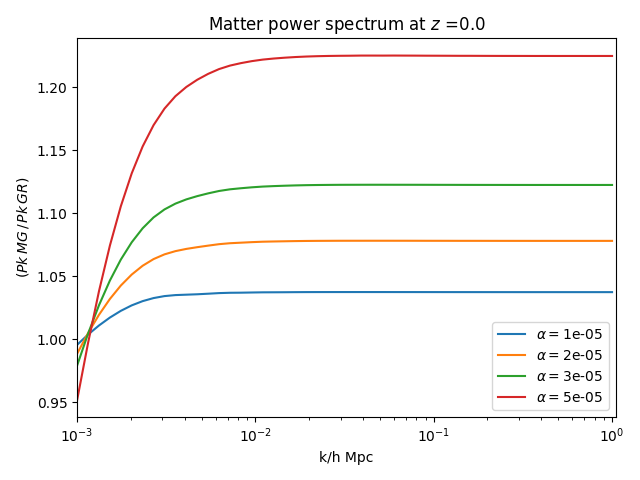}
    \includegraphics[width=0.35\textwidth, height=0.3\textwidth]{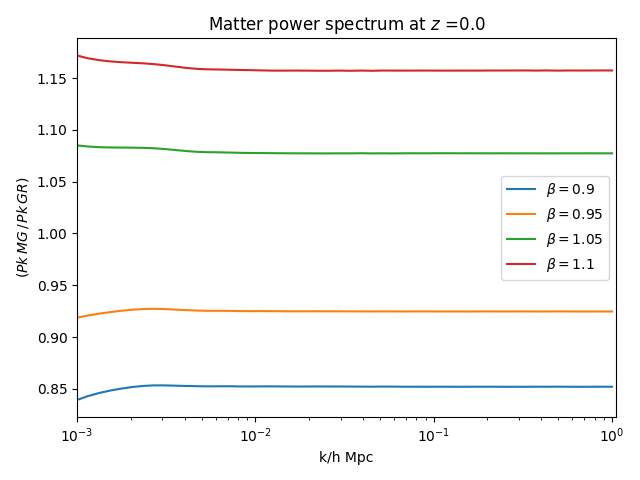}
    \includegraphics[width=0.35\textwidth, height=0.3\textwidth]{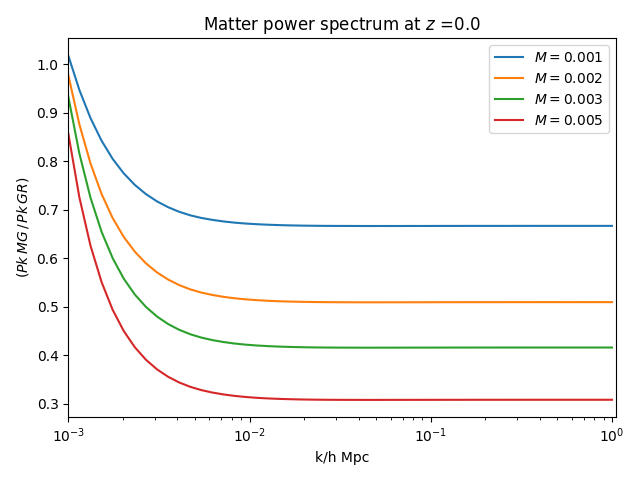}
    \caption[Power spectrum f(Q).]{Relative power spectrum of our model compared to GR for different values of $\alpha$, $\beta$ and $M$. }
    \label{fig:powerspectrum}
    \end{figure*}  
    
    \item We qualitatively tested the power spectrum outputted by the code. As it is a measure of clustering, it should decrease for a more rapid expanding background, i.e. higher $H(z)$, it should sink for smaller perturbation potentials, and on the opposite it should raise for smaller $H(z)$ and bigger perturbation potentials, which is seen in Fig.~\ref{fig:powerspectrum} and agrees with the results shown in \cite{Goncalves:2024sem}.    
\end{enumerate}
We use CMB temperature, polarization, their cross correlations C$_\ell$ and lensing spectrum D$_\ell$ likelihood \cite{Planck:2019nip} and data released by the Planck satellite mission \cite{Planck18.6} (Plk18) which we combine with background observations from BAO measurements \cite{Beutler:2011hx,Ross:2014qpa,BOSS:2016wmc}.  The constraints obtained from these two probes will agree, in $\Lambda$CDM with the high redshift values for $H_0$ and $\sigma_8$. To check whether our models will be able to reduce or solve the discrepancy, we infer the cosmological parameters within our $f(Q)$ theory using first the shear two-point correlation function likelihood and data from the fourth data release of the Kilo-Degree Survey (KiDS) together with the KiDS Cosmology Analysis Pipeline (KCAP) \cite{KiDS_ugriImaging, KiDS_methodology, KiDS_shearConstraints, KiDS_c1, KiDS_c2, KiDS_c3, KiDS_c4, KiDS_c5}. In this procedure the matter power spectrum gets projected along the line of sight to calculate the theoretical spectra $C_\mathrm{GG}$, $C_\mathrm{GI}$, and $C_\mathrm{II}$ which are then in turn converted into $\xi_{\pm}$ and compared to the measured values \cite{KiDS_shearConstraints}. We account for modified gravity by multiplying the integrands of $C_\mathrm{GG}$ and $C_\mathrm{GI}$ in the likelihood by a factor of $\mu^2$ and $\mu$ respectively. We change the integrands because, in analogy to \cite{Schmidt_2008}, the Poisson equation \ref{mod_poisson} is modified by a factor of $\mu$ while the equality of $\Psi$ and $\Phi$ still holds. Further we did check that the impact of the non-linear evolution of the matter power spectrum in $\Lambda$CDM on the $\chi^2$ statistic is below $0.1\%$ if we cut the scales for $\xi_+(\theta)$ and $\xi_-(\theta)$ below $\theta=\ang{;10;},\ang{;60;}$, respectively. We then follow the same procedure and also confront to observations, the theoretical predictions from the galaxy lensing, clustering and their cross correlated spectrum from the dark energy survey (DES) collaboration \cite{DES:2017myr,DES:2021wwk} in which we also limit to the linear scales. We run our MCMC using \texttt{MGCLASS II} \cite{MGCLASS2} \footnote{\url{https://gitlab.com/zizgitlab/mgclass--ii}} which is interfaced with the cosmological data analysis code \texttt{MontePython} \cite{MoPy3} in which the DES and KiDS public likelihoods were implemented and adapted by us for our submodels. We note that, in the case when we use DES, the $\beta$ parameter was showing multimodal behaviour with low probability for points that are far from the maximum likelihood. That is why we limit its space by a prior of 1\,$\sigma$ around its maximum likelihood obtained from earlier runs. At the end, to consolidate and break degeneracies when letting all our parameters free, we as well combine with redshift space distortion data (RSD) based on the dataset and likelihood by \cite{Sakr:2023bms}.

\section{Results and discussion}\label{sect:results}

\indent{\,\, \quad} Here we show the constraints on the two cosmological parameters subject to tensions, $H_0$ and $\sigma_8$, along with their related degenerate parameters, $\Omega_{m,0}$ and $S_8$. They were inferred following MCMC runs using CMB combined with BAO and then compared against those obtained from weak lensing shear correlations from KiDS survey. To this baseline we also show constraints from 3$\times$2pt joint analysis of photometric weak lensing and galaxy clustering alone or combined with growth measurements. We shall present different cases in which we allow each of our $f(Q)$ parameters to vary, before we consider different combinations of these parameters in addition to the aforementioned cosmological parameters.   

\begin{figure}
    \centering
    \includegraphics[width=0.9\linewidth]{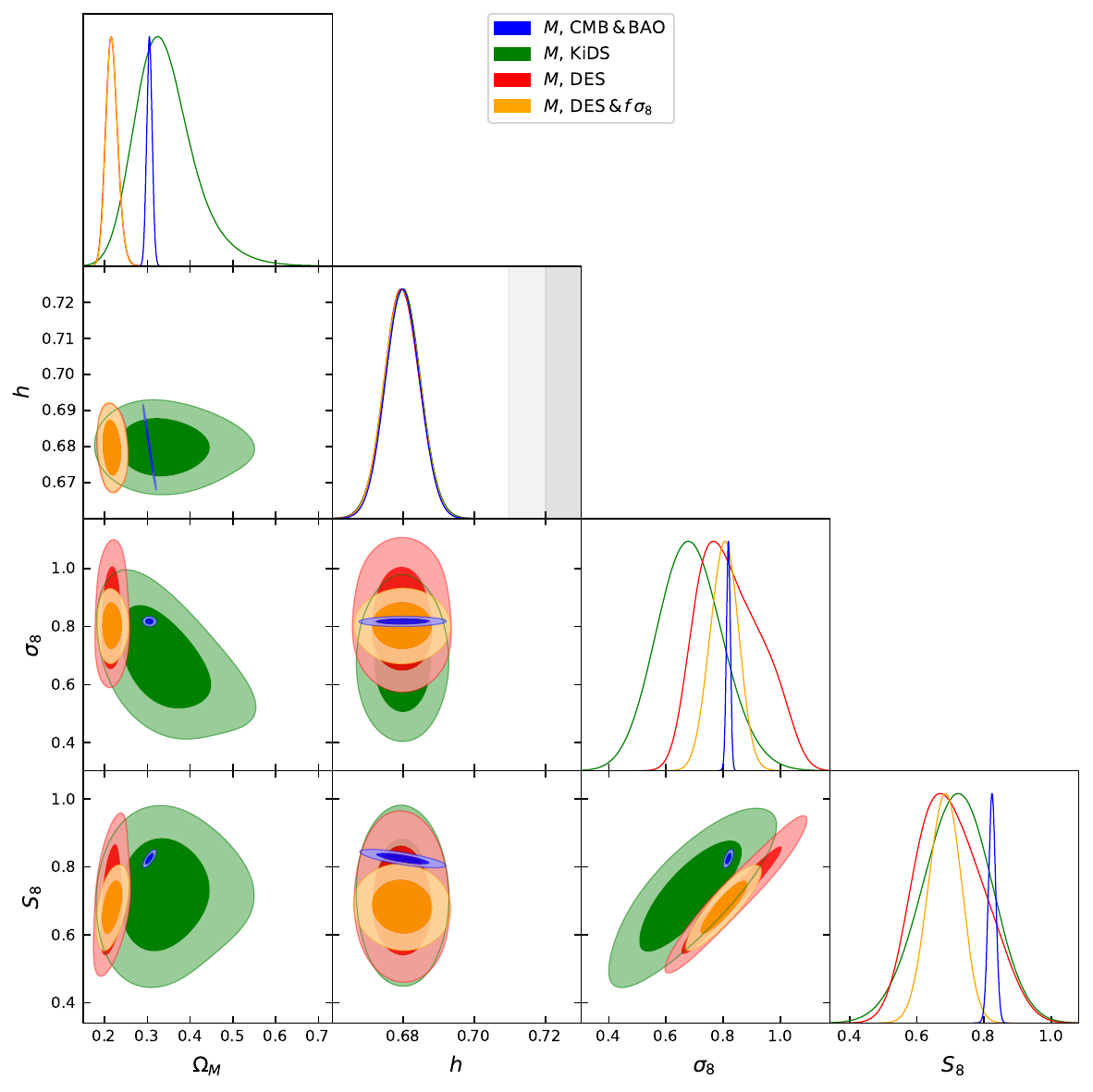}
    \caption{1D \& 2D 68\% and 95\% confidence contours within STG theory for $\Omega_{m,0}$, $h$, $\sigma_8$, $S_8$ and the $M$ free $f(Q)$ model parameter from the MCMC analysis using CMB combined with BAO vs Weak lensing shear correlations constraints vs 3$\times$2pt joint analysis of photometric weak lensing and galaxy clustering combined with $f\sigma_8$ from growth measurements. The gray (light gray) band represents respectively the 1$\sigma$ (2$\sigma$) errors on the value of inferred of $H_0$ from the SH0ES measurement.}
    \label{fig:M_CMBvsDESvsDES+fs8vsKids}
\end{figure}

We start in Fig.~\ref{fig:M_CMBvsDESvsDES+fs8vsKids} in which we first vary the $M$ parameter which is expected to have an impact on the $\sigma_8$ constraints alone since it theoretically only affects perturbations. Indeed, we observe no change to the Hubble parameter constraints, while either $S_8$ or the $\sigma_8$ tension is reduced to 1\,$\sigma$ when comparing with KiDS constraints. Although the latter are wider than those obtained from the 3$\times$2pt probe from DES as one would expect, the much tighter ones are still showing the same agreement on the $S_8$ level while the $\sigma_8$ is almost aligned with that coming from CMB+BAO. However the constraints on $\Omega_{m,0}$ have shrunk and show again a 2\,$\sigma$ discrepancy with that obtained from CMB+BAO. The addition of constraints from $f\sigma_8$ measurements confirm the DES findings on $\sigma_8$ - $\Omega_m$ but also limit the constraints on $S_8$ which then show again a 3\,$\sigma$ discrepancy with CMB+BAO. In comparison to related studies, which we can only do for this parameter since the other two remaining ones and their combinations are for the moment only explored by us, \cite{can_fq_challenge}, used the measurements of Plk18 \cite{Planck18.6}, to constrain the $M$ parameter and received $M=-0.64^{+0.64}_{-0.60}$ while we got a best-fit value of $M=-0.38^{+0.47}_{-0.40}$ using also CMB alone with free $M$, agreeing with their results within $1\sigma$. The tighter constraints could be based on the additional analysis of CMB lensing effects. Extracting the information from the CMB in condensed form from the three shift parameters \cite{CMB_shift_parameters}, \cite{DGPish} got $\Omega_\mathrm{Q}\equiv \alpha/(2\sqrt{6})=-0.10^{+0.13}_{- 0.11}$. Our CMB run employing ($\beta=1$, $\alpha\in\mathbb{R}$, $M=0$) delivered $\Omega_\mathrm{Q}=0.011\pm{+0.015}$, within the $68\%$ confidence region of their best-value but tighter due to the usage of the whole TT, TE, EE spectra and the lensing likelihood. Finally, in \cite{RSD_fq_paper}, the authors received $M=2.0331^{+3.8212}_{-1.9596}$ from redshift space distortion measurements while we get, when we limit to RSD data, $M=2.37^{+5.62}_{-2.94}$, in 1-$\sigma$ agreement with them. 

\begin{figure}
    \centering
    \includegraphics[width=0.9\linewidth]{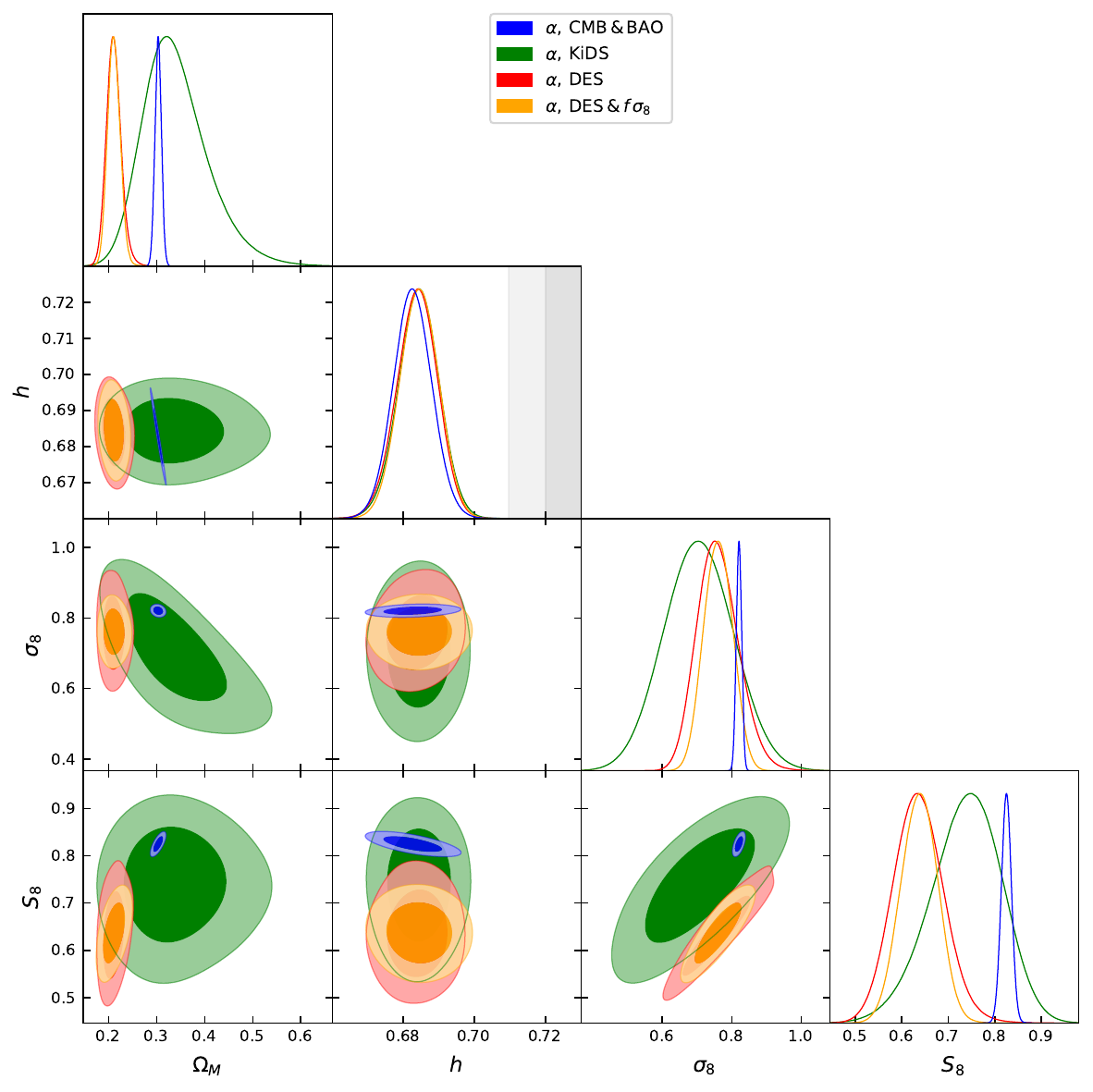}
    \caption{1D \& 2D 68\% and 95\% confidence contours within STG theory for $\Omega_{m,0}$, $h$, $\sigma_8$, $S_8$ and the $\alpha$ free $f(Q)$ model parameter from the MCMC analysis using CMB combined with BAO vs Weak lensing shear correlations constraints vs 3$\times$2pt joint analysis of photometric weak lensing and galaxy clustering combined with $f\sigma_8$ from growth measurements. The gray (light gray) band represents respectively the 1$\sigma$ (2$\sigma$) errors on the value of inferred of $H_0$ from the SH0ES measurement.}
    \label{fig:alphaCMBvsDESvsDES+fs8vsKids}
\end{figure}

Varying next the $\alpha$ parameter alone in Fig.~\ref{fig:alphaCMBvsDESvsDES+fs8vsKids}, we first observe, for each of the probe combinations, a small change in the $H_0$ parameter constraints, widening its 3\ $\sigma$ limit slightly beyond $\sim$ 70.0, marginally reducing the discrepancy with $H_0 \sim$ 73.5 obtained from SH0ES. The $\sigma_8$ tension is similarly to the $M$ case reduced when using KiDS, while the more constraining 3$\times$2pt probe from DES, with or without $f\sigma_8$, are limiting $S_8$ or the $\Omega_m$ to values in more than 3\,$\sigma$ from those inferred from CMB+BAO, despite an elevation of the $\sigma_8$ tension when considered alone.
 
\begin{figure}
    \centering
    \includegraphics[width=0.9\linewidth]{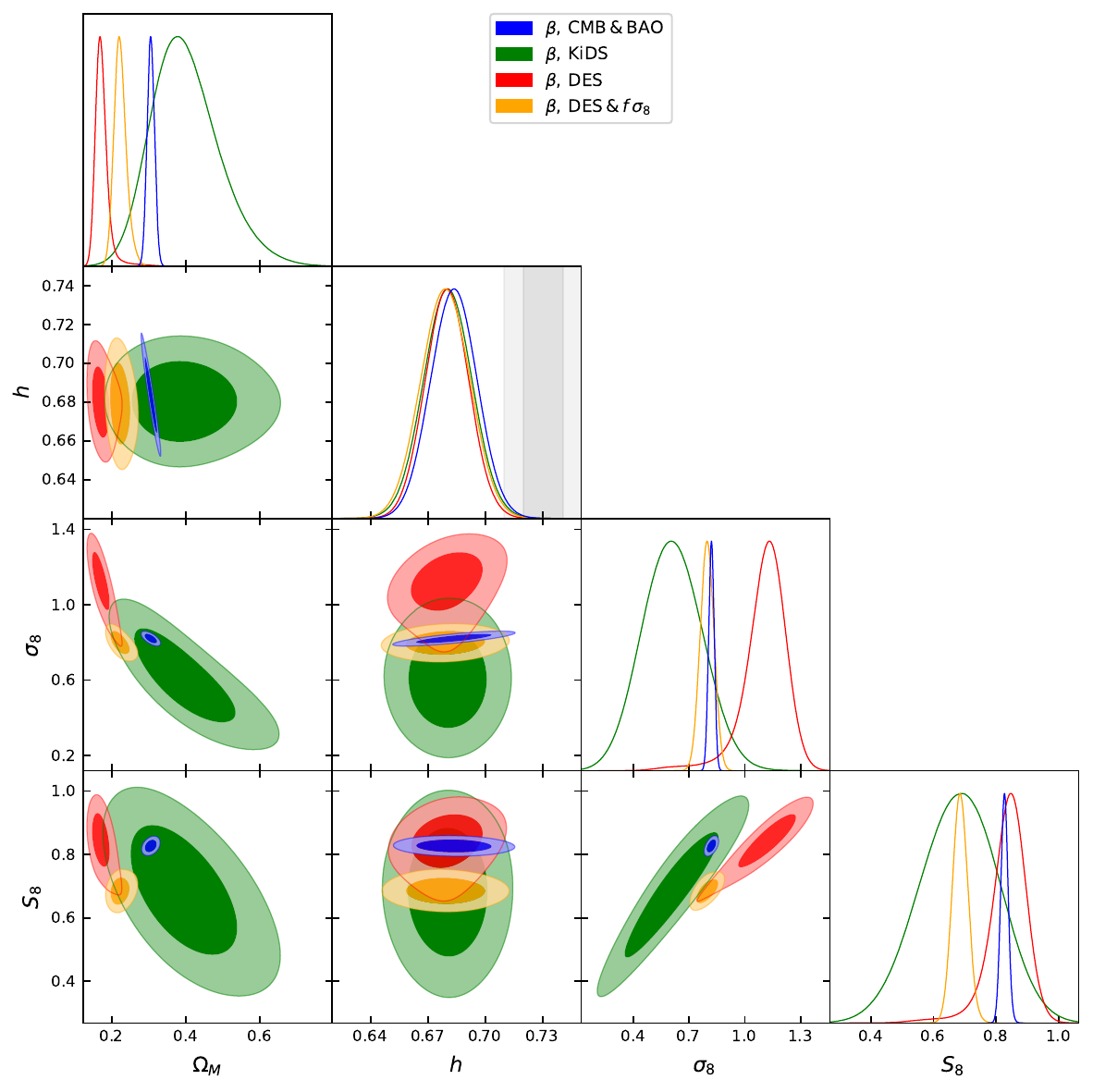}
    \caption{1D \& 2D 68\% and 95\% confidence contours within STG theory for $\Omega_{m,0}$, $h$, $\sigma_8$, $S_8$ and the $\beta$ free $f(Q)$ model parameter from the MCMC analysis using CMB combined with BAO vs Weak lensing shear correlations constraints vs 3$\times$2pt joint analysis of photometric weak lensing and galaxy clustering combined  with $f\sigma_8$ from growth measurements. The gray (light gray) band represents respectively the 1$\sigma$ (2$\sigma$) errors on the value of inferred of $H_0$ from the SH0ES measurement.}
    \label{fig:betaCMBvsDESvsDES+fs8vsKids}
\end{figure}

We end with constraints obtained when still considering only one free $f(Q)$ parameter $\beta$ shown in Fig.~\ref{fig:betaCMBvsDESvsDES+fs8vsKids}. The $H_0$ 95\% confidence contours extend to $\sim$ 72.0, substantially reducing the Hubble tension and that for all our probes, while the $\sigma_8$ tension is alleviated when constrained by KiDS, albeit a widening in the constraints with respect to the tight ones obtained from CMB+BAO probe. This is not the case for when we constrain with 3$\times$2pt from DES despite the fact that the $S_8$ maximum 1D likelihood contours overlap with that from the CMB+BAO combination since now the $\Omega_{m,0}$ - $\sigma_8$ is showing discrepancy. Combining with $f\sigma_8$ further degrade the situation where we end by a strong discrepancy on the $\Omega_{m,0}$ - $\sigma_8$ as well as on the $S_8$ parameters.   

\begin{figure}
    \centering
    \includegraphics[width=0.9\linewidth]{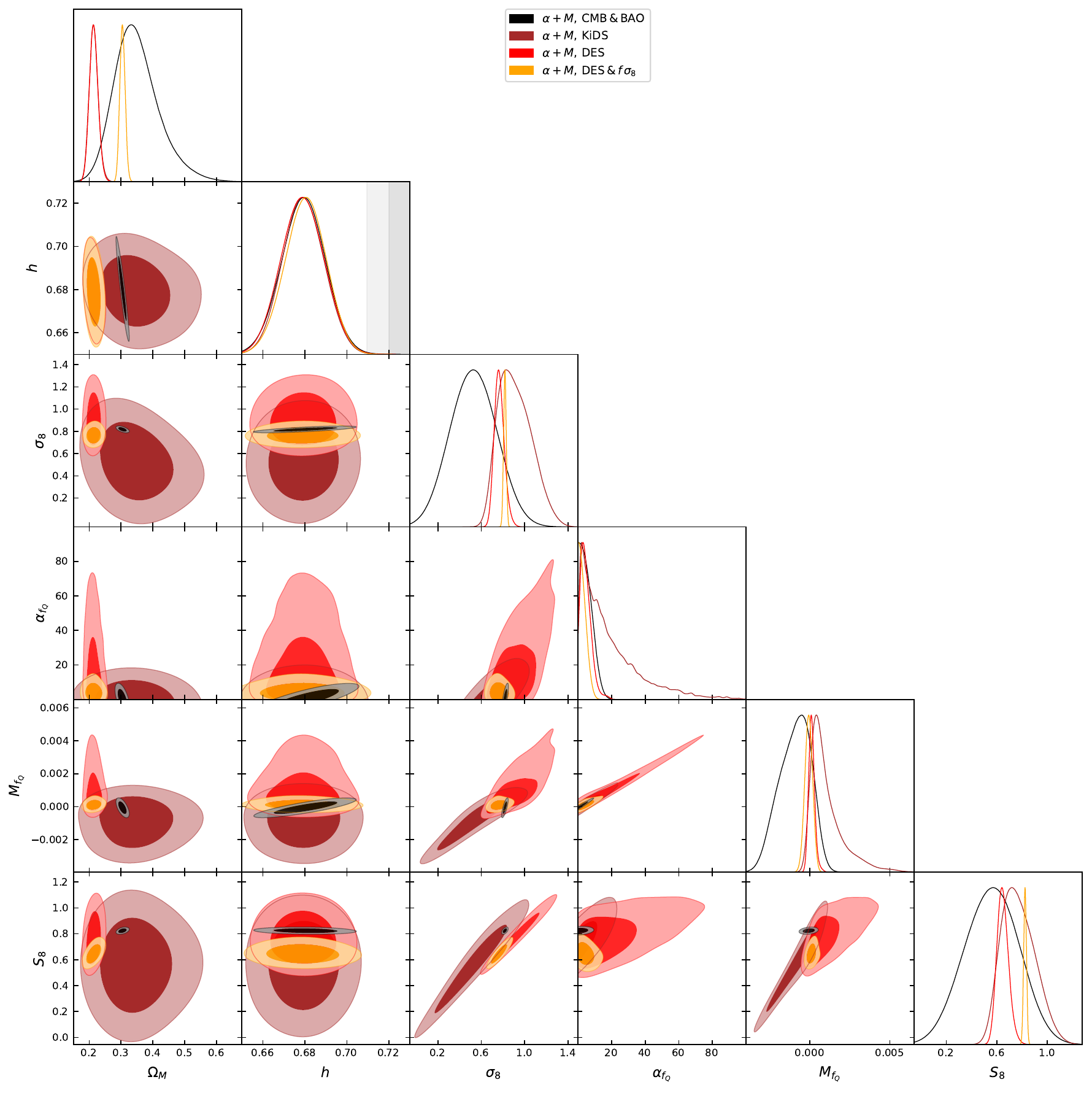}
    \caption{1D \& 2D 68\% and 95\% confidence contours within STG theory for $\Omega_{m,0}$, $h$, $\sigma_8$, $S_8$ and the $\alpha$ and $M$ free $f(Q)$ model parameters from the MCMC analysis using CMB combined with BAO vs Weak lensing shear correlations constraints vs 3$\times$2pt joint analysis of photometric weak lensing and galaxy clustering combined  with $f\sigma_8$ from growth measurements. The gray (light gray) band represents respectively the 1$\sigma$ (2$\sigma$) errors on the value of inferred of $H_0$ from the SH0ES measurement.}
    \label{fig:alpha+M_CMBvsDESvsDES+fs8vsKids}
\end{figure}

With no single parameter being able of totally alleviating our two considered tensions, we allow next more than one free $f(Q)$ parameter at once, starting by combining with the $M$ parameter that only affect the perturbations, to move next to considering the $\alpha$ \& $\beta$ parameters before ending by allowing more freedom when we let our three $f(Q)$ parameterization parameters to vary at once. Therefore, we show in Fig.~\ref{fig:alpha+M_CMBvsDESvsDES+fs8vsKids} confidence contours for when letting free $\alpha$ and $M$. The Hubble constant is still marginally reduced, the same as $\alpha$ alone case as expected, and we see the same constraints on $\sigma_8$ when using KiDS, while the combination of $\alpha$ and $M$ is allowing the constraints from  3$\times$2pt to also reduce the tension, albeit it is restored again when we combined with 
$f\sigma_8$. We end by noting that we restricted $\alpha$ to positive values which explains on why its 1D contour seems smaller in the case of 3$\times$2pt with respect to when we further add $f\sigma_8$, while the large parameter space expands in the negative values.

\begin{figure}
    \centering
    \includegraphics[width=0.9\linewidth]{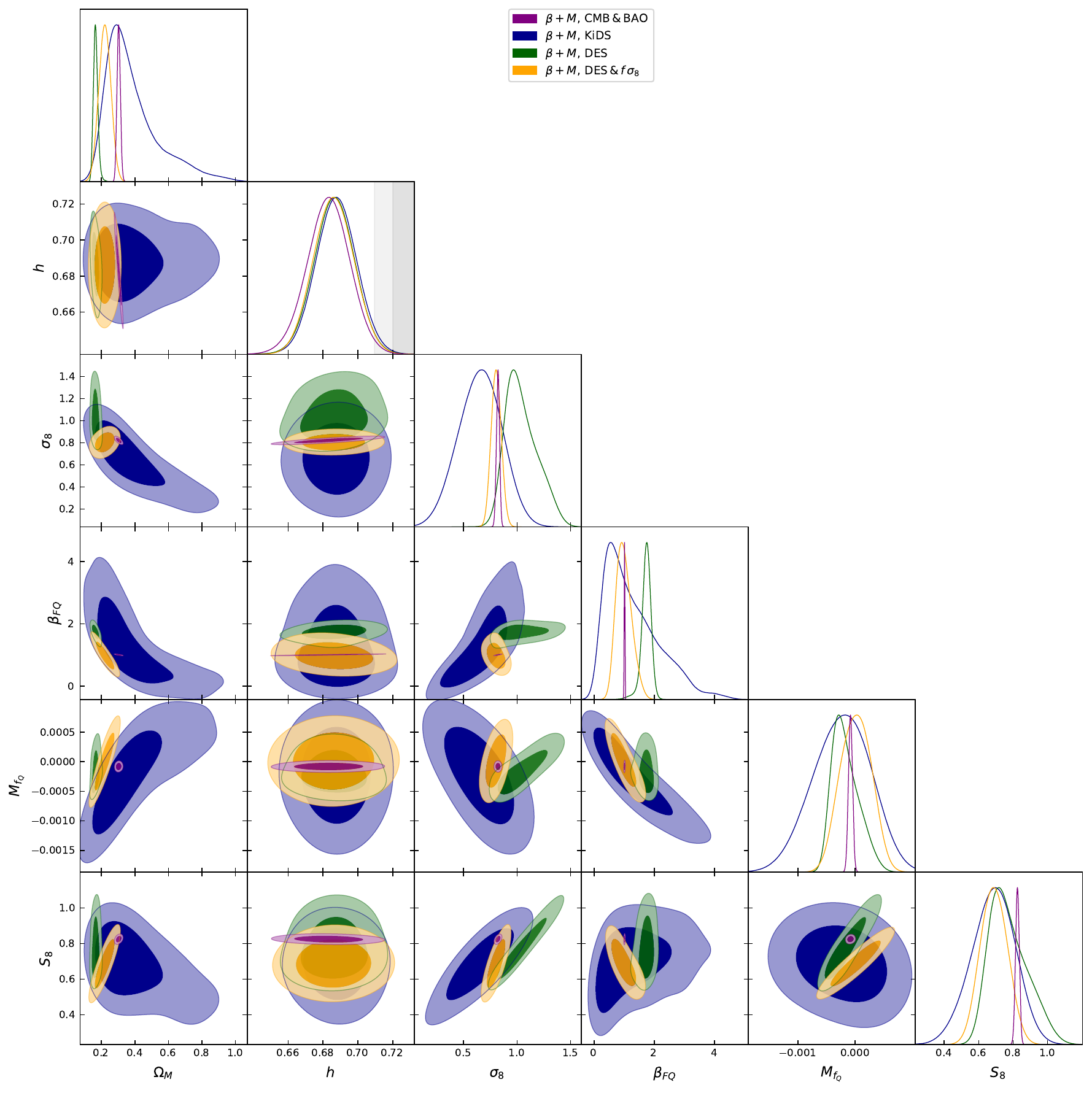}
    \caption{1D \& 2D 68\% and 95\% confidence contours within STG theory for $\Omega_{m,0}$, $h$, $\sigma_8$, $S_8$ and the $\beta$ and $M$ free $f(Q)$ model parameters from the MCMC analysis using CMB combined with BAO vs Weak lensing shear correlations constraints vs 3$\times$2pt joint analysis of photometric weak lensing and galaxy clustering combined  with $f\sigma_8$ from growth measurements. The gray (light gray) band represents respectively the 1$\sigma$ (2$\sigma$) errors on the value of inferred of $H_0$ from the SH0ES measurement.}
    \label{fig:beta+M_CMBvsDESvsDES+fs8vsKids}
\end{figure}

\begin{figure}
    \centering
    \includegraphics[width=0.9\linewidth]{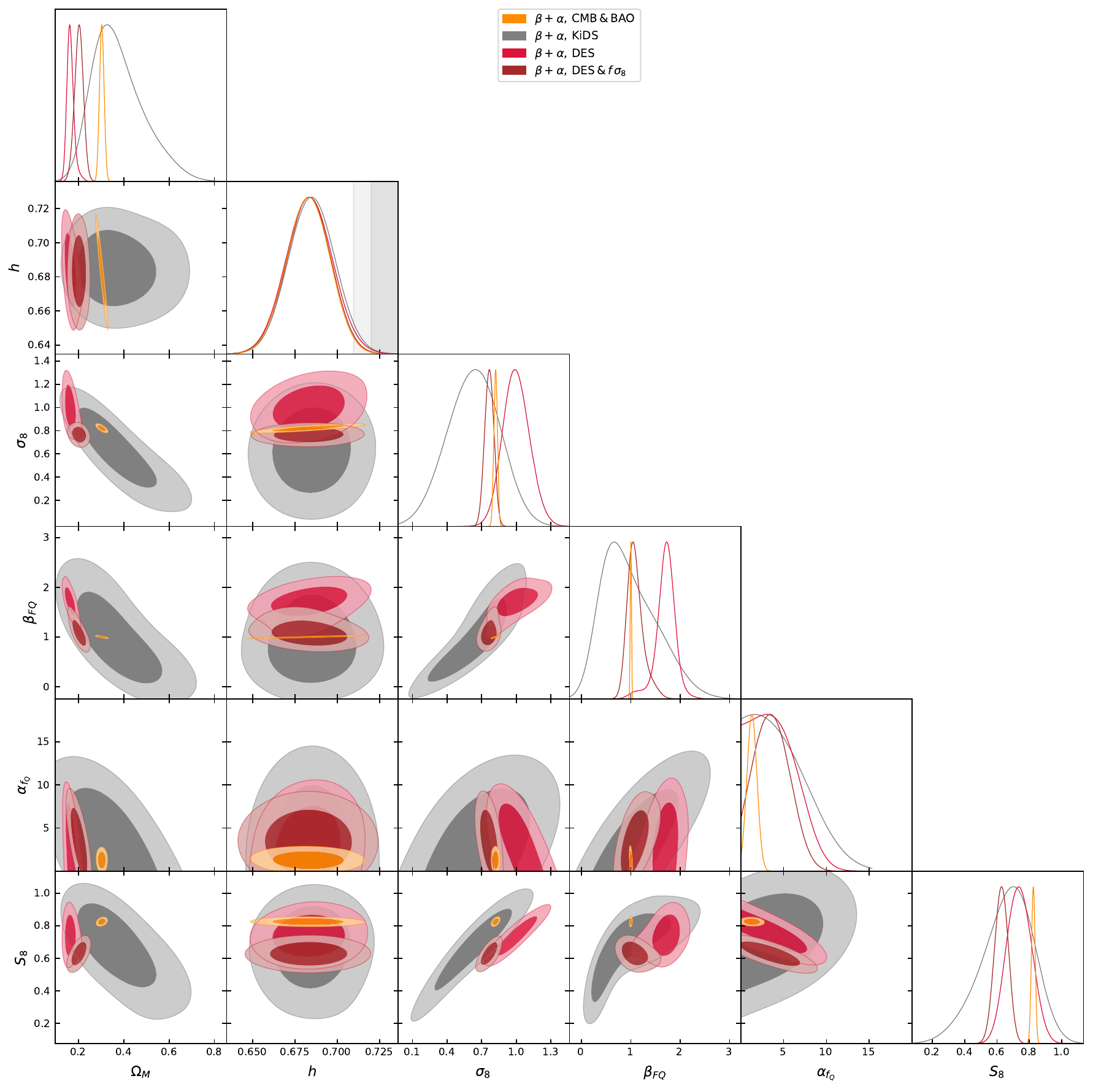}
    \caption{1D \& 2D 68\% and 95\% confidence contours within STG theory for $\Omega_{m,0}$, $h$, $\sigma_8$, $S_8$ and the $\alpha$ and $\beta$ free $f(Q)$ model parameters from the MCMC analysis using CMB combined with BAO vs Weak lensing shear correlations constraints vs 3$\times$2pt joint analysis of photometric weak lensing and galaxy clustering combined  with $f\sigma_8$ from growth measurements. The gray (light gray) band represents respectively the 1$\sigma$ (2$\sigma$) errors on the value of inferred of $H_0$ from the SH0ES measurement.}
    \label{fig:alpha+beta_CMBvsDESvsDES+fs8vsKids}
\end{figure}

Allowing the parameter $M$ to vary along with $\beta$ as seen in Fig.~\ref{fig:beta+M_CMBvsDESvsDES+fs8vsKids} yields similar improvement on the ability to solve the tensions with moreover an $M$ - $\beta$ correlation observed in the most constraining case where 3$\times$2pt is combined with $f\sigma_8$, allowing a reduction to 1\,$\sigma$ for the $S_8$ parameter and less than 2\,$\sigma$ for the $\Omega_{m,0}$ - $\sigma_8$ confidence contours, while the combination of $\alpha$ and $\beta$, two parameters that affect the background evolution as well as the growth of perturbations, do not improve as much on the $\sigma_8$ parameter level as seen in Fig.~\ref{fig:alpha+beta_CMBvsDESvsDES+fs8vsKids} but rather extends the bounds on $H_0$ till values at 3\,$\sigma$ from the local ones for the Hubble parameter.

\begin{figure}
    \centering
    \includegraphics[width=0.9\linewidth]{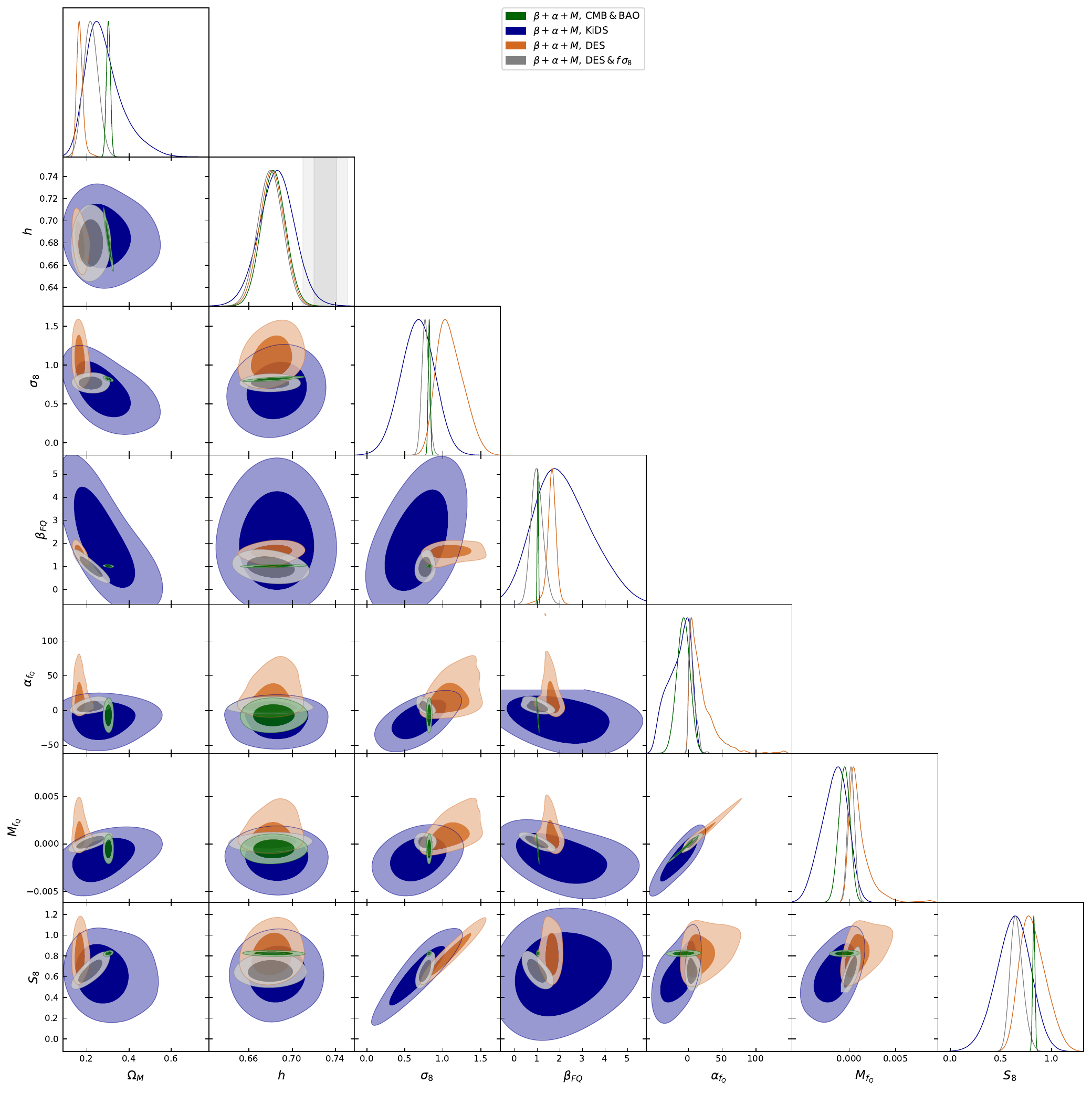}
    \caption{1D \& 2D 68\% and 95\% confidence contours within STG theory for $\Omega_{m,0}$, $h$, $\sigma_8$, $S_8$ and the $\alpha$, $\beta$ and $M$ free $f(Q)$ model parameters from the MCMC analysis using CMB combined with BAO vs Weak lensing shear correlations constraints vs 3$\times$2pt joint analysis of photometric weak lensing and galaxy clustering combined  with $f\sigma_8$ from growth measurements. The gray (light gray) band represents respectively the 1$\sigma$ (2$\sigma$) errors on the value of inferred of $H_0$ from the SH0ES measurement.}
    \label{fig:alpha+beta+M_CMBvsDESvsDES+fs8vsKids}
\end{figure}

We end by showing the case where we allow all degrees of freedom in our $f(Q)$ parameterisation as in Fig.~\ref{fig:alpha+beta+M_CMBvsDESvsDES+fs8vsKids} where $\alpha$, $\beta$ and $M$ are left free to vary. The Hubble tension is now within 2\,$\sigma$, even in the most constraining case while the $\sigma_8$ is within 1\,$\sigma$. We note that all the parameters are compatible with their $\Lambda$CDM values. We also note that we allowed $\alpha$ to explore negative values in order to further show that the multidimensional parameter space shrinks when we pass from the less constraining weak lensing shear from KiDS to the 3$\times$2pt from DES. 

 \section{Conclusion}\label{sect:conclusion}

The era of precision cosmology has revealed tensions between predictions of {$\Lambda$CDM} and increasingly accurate observations. These discrepancies might be overcome by a modification of standard General Relativity, such as the rapidly developing Symmetric Teleparallel non-metric compatible Gravity in which the Ricci scalar in the action is replaced by a general function $f(Q)$ of the non-metricity scalar, Q. In this study, we incorporated a three-parameter family of $f(Q)$ models into the Boltzmann code \texttt{MGCLASS} at both the background and perturbation levels. 

We conducted a Bayesian study by means of Markov Chain Monte Carlo methods, varying along with the cosmological parameters, three others parameterizing $f(Q)$, namely $M$ impacting only on the perturbation level, $\alpha$ modifying the Friedmann equation by scaling an additional Hubble parameter and $\beta$ which further scales the Gravitational constant. Our analysis focused on the impact of the Hubble tension in $H_0$ and the discrepancy in $\sigma_8$ resulting from the inclusion of our $f(Q)$ model’s parameters. We employed probes that included the cosmic microwave background (CMB) temperature, polarization and lensing power spectrum from Planck 2018 datasets, baryon acoustic oscillations (BAO) from eBOSS survey, weak lensing (WL) from KiDS survey, 3x2pt lensing and galaxy clustering tomographic correlations from DES survey, and growth measurements from the RSD effect.  

We find that none of the sub models, considered alone, were able of substantially reducing the Hubble tension with the strongest impact coming from varying the $\beta$ parameter bringing the discrepancy to the 3\,$\sigma$ level. When combined, we found that, at best, in the highest degree of freedom case, were all the $f(Q)$ parameters are left free, the Hubble is further reduced to the 2\,$\sigma$ level. On the other hand, we found that the $S_8$ discrepancy between CMB+BAO versus KiDS bounds, already mitigated on WL linear scales at which we choose to cut our KiDS survey data, is also alleviated when using the 3x2pt probe from DES, especially when we let all our model's parameters as free with the $\beta$ parameter in particular showing preference for values far from the GR ones. However, the $\Omega_{m,0}$ - $\sigma_8$ discrepancy remained unchanged. When considering combinations with more constraining probes, such as structure clustering $f\sigma_8$ growth measurements, $\beta$ was constrained back to being compatible with GR but the $S_8$ discrepancy was not alleviated anymore. We conclude that the $f(Q)$ model we considered in this work, has the ability to reduce both the $H_0$ and $\sigma_8$ tension at the same time, without fully alleviating it, but the final call is awaiting more data coming from future surveys in order to break degeneracies and limit the space of variation of our model's parameters.

\acknowledgments

ZS acknowledges support from DFG project 456622116. ZS and LS thank Luca Amendola for useful discussions on this topic.

\bibliographystyle{JHEP}
\bibliography{JHEP.bib}

\end{document}